\documentclass[pra,twocolumn,superscriptaddress,showpacs,nofootinbibfloatfix,amsmath,amsfonts,amssymb]{revtex4-1}%
\usepackage{braket}
\usepackage{tabularx}

\usepackage{dsfont}
\usepackage{amsmath,amsfonts,amssymb}
\usepackage{float}
\usepackage{graphicx}
\usepackage{dcolumn}
\usepackage{bm}
\usepackage{hyperref}

\usepackage{amsmath,amsfonts,amssymb,color}
\usepackage{amsthm}
\usepackage{leftidx}
\usepackage{graphicx}
\usepackage{xcolor}
\usepackage{dcolumn}
\usepackage{bm}
\usepackage{epstopdf}
\usepackage{epsfig}
\usepackage{environ}
\usepackage{pdfcomment}

\usepackage{float}
\usepackage[T1]{fontenc}
\usepackage[latin9]{inputenc}
\usepackage{setspace}
\usepackage{esint}

\begin{document}

\preprint{APS/123-QED}

\title{{Floquet band engineering with Bloch oscillations}}





\author{Xi Liu}
\affiliation{NUS Graduate School - Integrative Sciences and Engineering Programme (ISEP), National University of Singapore, Singapore 119077, Singapore}

\author{Senmao Tan}
\affiliation{Department of Physics, National University of Singapore, Singapore 117551, Singapore}

\author{Qing-hai Wang}
\affiliation{Department of Physics, National University of Singapore, Singapore 117551, Singapore}

\author{Longwen Zhou}
\email{zhoulw13@u.nus.edu}
\affiliation{College of Physics and Optoelectronic Engineering, Ocean University of China, Qingdao 266100, China}

\author{Jiangbin Gong}
\email{phygj@nus.edu.sg}
\affiliation{Department of Physics, National University of Singapore, Singapore 117551, Singapore}
\affiliation{Center for Quantum Technologies, National University of Singapore, Singapore 117543, Singapore}
\date{\today}

\begin{abstract}
 {This work provides a convenient and powerful means  towards the engineering of Floquet bands via Bloch oscillations, by adding a tilted linear potential to periodically driven lattice systems. The added linear field not only restricts the spreading of a time-evolving wavepacket but also, depending on the ratio between the Bloch oscillation frequency and the modulation frequency of the periodic driving, dramatically modifies the band profile and topology.
Specifically, we consider a driven Aubry-Andr\'{e}-Harper model as a working example, in the presence of a linear field.  Almost flat Floquet bands  or Floquet bands with large Chern numbers due to the interplay between the periodic driving and Bloch oscillations can be obtained, with the band structure and topology extensively tunable by adjusting the ratio of two competing frequencies. To confirm our finding,   we further execute the Thouless pumping of one and two interacting bosons in such a lattice system and establish its connection with the topological properties of single- and two-particle Floquet bands.  }
\end{abstract}

\pacs{}
\keywords{}
\maketitle
\section{Introduction}\label{Sec:int}

Topological Thouless pumping describes the quantized adiabatic transport of particles in a one-dimensional (1D) lattice, which is induced by the slow and time-periodic variation of the lattice potential~\cite{thoulessPump}. It reveals the deep connection among geometric phase, band topology and quantum adiabatic process~\cite{thouless1982quantized,niu1984quantised,king1993theory,XiaoRMP2010,Gritsev2012}. The observations of Thouless pumping in setups like cold atoms and photonics not only demonstrate the topology of integer quantum Hall effect from a dynamical perspective, but also provides a powerful tool for the detection of topological invariants through quantum dynamics~\cite{photonicpumpingexperiment,thoulessPumpUltracoldAtomsOpticalSupperlattice,topologicalThoulessPumpUltracoldFermion,TPExp4,TPExp5}. 

In the conventional Thouless pump, the applied adiabatic driving field plays the role of a second synthetic dimension~\cite{XiaoRMP2010}. To guarantee the pumping quantization, the initial state of the system needs to be a Wannier state or a uniformly filled Bloch band along the physical dimension, which can be rather challenging to prepare in experiments~\cite{thoulessPumpUltracoldAtomsOpticalSupperlattice,topologicalThoulessPumpUltracoldFermion}. Recently, Bloch oscillations induced by a linearly tilted potential were found to be able to facilitate the initial state preparation in Thouless pump~\cite{topologicalPumpingAssisted}. Specially, the Bloch oscillations assist in the uniform sampling of eigenstates at all quasimomenta in the first Brillouin zone, so that nearly perfect pumping quantization can be achieved for an arbitrary initial state selected in a band of interest. Adding more interest to studies of Bloch oscillations versus topological phases, Bloch oscillations have also been used in experiments to probe the band topology~\cite{anABInterferometer, measurinTheChernNumberOfHofstadter,directMeasurementOfTheZak}. For systems with many-body interactions, the introduction of a linear potential is also of great interest, because a time-evolving state can be localized by such a linear potential due to the so-called Stark many-body localization~\cite{starkManybody}.

Beyond the adiabatic limit, time-periodic driving fields were found to induce rich Floquet topological phases~\cite{OkaPRB2009,floquetQuantumWells,ho2012quantized,floquetTopologicalInsulators,GrushinPRL2014,TitumPRL2015,LeykamPRL2016,ZhouPRB2016,periodicTabelForFloquet,PhysRevLett.120.230405,PhysRevLett.121.036402,YapPRB2018,ZhouPRA2018,ZhouPRB2018,floquetDynamicalQuantumPhaseTransitions,PhysRevB.99.045441,bandStructureEngineering,AFINanophotonic,AFTUltracoldAtoms,lightInducedAHE,
tan2020high,ZhouPRA2020,PhysRevB.101.085401,
PhysRevResearch.3.L032026,PhysRevB.104.L020302,topologicalSpinTexture}. An extension of the Thouless pump to Floquet systems was introduced as a powerful dynamical framework to probe the large Chern number phases of Floquet topological insulators~\cite{ho2012quantized,aspectsOf,xiong2016towards}. However, the Floquet-Thouless pump defined in Ref.~\cite{aspectsOf}  requires the system to be prepared in a localized Wannier state or as a uniformly populated Floquet-Bloch band, which is even more challenging to achieve in experiments due to the lack of a Fermi surface in periodically driven systems~\cite{ibctheory,Raghava2017,ibcexperiment}. One may then consider adding a linear potential to a Floquet system, and await the resulting Bloch oscillations to assist the initial state preparation in a Floquet Chern band and thus yielding the quantized Thouless pumping. The actual situation is, however, more subtle and complicated. Indeed, in Floquet systems,  once a linear field is introduced, the resulting effective Hamiltonian of the system associated with one whole period of driving is NOT the original effective Hamiltonian plus the linear field under considetation. Qualitatively, this is because of the interplay between the linear field and the microscopic motion within one driving period. Specifically, let $H[\theta_{1}(t)]+F=H[\theta_{1}(t+2\pi/\Omega)]+F$ be the time-periodic Hamiltonian of a tight-binding lattice under a linearly tilted potential $F$, where $T_{1}=2\pi/\Omega$ is the Floquet driving period and $\Omega$ is the driving frequency. By transforming $H[\theta_{1}(t)]+F$ into a rotating frame, the linear potential $F$ can be replaced by an oscillating phase factor $\theta_{2}(t)=\omega_{F} t$ in the hopping amplitude of the lattice, where $T_{2}=2\pi/\omega_{F}$ is the oscillation period and $\omega_{F}$ is the field strength of the linear potential. The rotated Hamiltonian $H_{\rm r}[\theta_{1}(t),\theta_{2}(t)]$ now recovers the translational symmetry along the physical dimension, but as a cost carries two separate time-dependencies through the parameters $\theta_{1}(t)$ and $\theta_{2}(t)$. If the ratio between the two driving frequencies $\Omega/\omega_{F}$ is irrational, the rotated Hamiltonian can no longer describe a Floquet system. When the frequency ratio $\Omega/\omega_{F}\in{\mathbb Q}$, $H_{\rm r}[\theta_{1}(t),\theta_{2}(t)]$ could still describe a Floquet model, but with a new driving period $T$ that is equal to the least common multiple of $T_{1}$ and $T_{2}$. Nevertheless, the two driving fields may couple and interplay strongly in this case, yielding Floquet topological bands and Thouless pumping that are totally different from those that one could expect in the system described by $H[\theta_{1}(t)]$. Therefore, the impacts of a linearly tilted potential (and its resulting Bloch oscillations) on the topology, initial state preparation and adiabatic transport in Floquet systems are highly nontrivial and deserve to be explored in detail.

{In this work, we uncover how the Bloch oscillations induced by a linear potential explicitly modify the topology of quasienergy bands and control the Thouless pumping in a Floquet system. We find that the added linear field could not only restrict the spreading of a time-evolving wavepacket, but also dramatically modifies the profile and topology of Floquet bands when the Floquet driving frequency and linear field strength are set at appropriate rational ratios. Focusing on a driven Aubry-Andr\'{e}-Harper model in the presence of a linear field as a working example, we find almost flat Floquet bands or Floquet bands with large Chern numbers due to the interplay between the periodic driving and Bloch oscillations. The band structures and topology are further found to be highly tunable by adjusting the ratio of two competing frequencies. To confirm our findings, we further execute the Thouless pumpings of one and two interacting bosons in the system, and establish their connections with the topological properties of single- and two-particle Floquet quasienergy bands.}

The rest of the manuscript is structured as follows. 
In Sec.~\ref{Sec:the}, we introduce a theoretical framework to describe the band structure and dynamics of bosons in a 1D superlattice subject to time-periodic driving fields, particle-particle interactions and an onsite linear potential. 
In Secs.~\ref{Sec:boson1} and \ref{Sec:boson2}, we apply our theory to reveal the Floquet band topology and quantized Thouless pumping in one- and two-particle systems with an emphasis on the role played by the linear potential and the resulting Bloch oscillations. 
In Sec.~\ref{Sec:sum}, we summarize our results and discuss potential future directions.
Some further calculation details and results are presented in the Appendices \ref{derivation}--\ref{moreExamples}.

\section{Theory}\label{Sec:the}
Here we introduce the system that we are going to explore and outline the theoretical framework that will be used to study its spectral, topological and transport features.

\subsection{Hamiltonian and pumping dynamics}\label{Sec:H}
\label{hamiltonian}
We start with a harmonically driven AAH model~\cite{aspectsOf} plus a linear onsite potential and an interaction term. The resulting Hamiltonian reads
\begin{align}
    H&=\sum_j\left[\frac{J}{2}(\hat{a}_{j}^{\dagger}\hat{a}_{j+1}+{\rm H.c.})+V\cos(\lambda j-\beta)\cos(\Omega t)\hat{n}_{j}\right]\nonumber\\
    {}&+\sum_{j}\frac{U}{2}\hat{n}_{j}(\hat{n}_{j}-1)+\sum_{j}\omega_{F}j\hat{n}_{j}.\label{realHamiltonian}
\end{align}
Here $\hat{a}_{j}^{\dagger}$ ($\hat{a}_j$) creates (annihilates) a boson on the sublattice site $j$. $J$ is the hopping amplitude. $V$ is the driving strength. $\Omega$ is the driving frequency. $\omega_{F}$ is the strength of linear potential. $\lambda=2\pi p/q$, with $p$ and $q$ being coprime integers. We choose $p/q=1/3$ in this work, so that the spectrum of $H$ possesses three bands in the absence of the linear potential and many-body interactions. $\beta\in[0,2\pi]$ is a phase shift of the superlattice potential, which will be tuned adiabatically in the Floquet Thouless pumping. More precisely, we consider the evolution over a time duration $MT$ with $M\gg1$ and $T$ the driving period in the pumping dynamics. Within the $m$th driving period ($m=1,2,...,M-1,M$), we let $\beta=2\pi (m-1)/M$ and evolve the state using the Hamiltonian in Eq.~(\ref{realHamiltonian}) from $t=(m-1)T$ to $t=mT$. The pumping is revealed by the expectation value of unit-cell position operator $\hat{x}=\frac{1}{qN}\sum_{j}j\hat{n}_{j}$ at the end of each driving period, with $N$ the particle number.

The linear potential in Eq.~(\ref{realHamiltonian}) breaks the translational symmetry of $H$. Instead of directly studying the Hamiltonian in the lab frame, we apply a rotation to eliminate the linear potential term, so that the system can be made translationally invariant in real space \cite{EckardtRMP2017}. The rotation is defined as
\begin{equation}
    R=\sum_{\ket{\mathbf{n}}}e^{i\omega_{F}t\sum_{j}j{ n}_{j}}\ket{\mathbf{n}}\bra{\mathbf{n}},
\end{equation}
following which we get the Hamiltonian in the rotating frame, i.e.,
\begin{align}
    H_{r}&=\sum_{j}\frac{J}{2}(e^{-i\omega_{F}t}\hat{a}_{j}^{\dagger}\hat{a}_{j+1}+{\rm H.c.})\nonumber\\
    {}&+\sum_{j}\left[V\cos(\lambda j-\beta)\cos(\Omega t)\hat{n}_{j}+\frac{U}{2}\hat{n}_{j}(\hat{n}_{j}-1)\right].\label{rotated}
\end{align}
As shown in Eq.~(\ref{rotated}), $H_{r}$ is subject to drivings with two frequencies $\omega_{F}$ and $\Omega$. In order to apply the Floquet formalism, we assume that $\omega_{F}$ and $\Omega$ are commensurate, i.e., letting their ratio be a rational number,
\begin{equation}
    \omega_{F}/\Omega=a/b,
\end{equation}
where $a,b$ are coprime integers. Let $T_{1}={2\pi}/\Omega$ and $T_{2}=2\pi/\omega_{F}$. The smallest common period of the two driving fields is then
\begin{equation}
    T=aT_{2}=bT_{1}.
\end{equation}
Let $\mathcal{T}$ be the time ordering operator, Floquet operator of the system at each fixed $\beta$ reads
\begin{equation}
    U(\beta)=\mathcal{T}\exp\left[-i\int_{0}^{T}dt H_{r}(t,\beta)\right].
\end{equation}
$U(\beta)$ governs the dynamics of the system over each common driving period of the two-color field. In the next subsection, we discuss how to use the translational invariance of the system over a unit cell to block diagonalize $U(\beta)$ in such  a system under the periodic boundary condition (PBC).

\subsection{Multiparticle Floquet-Bloch bands}
\label{diagonalization}
We now generalize the approaches of Refs.~\cite{aspectsOf,multiparticleWannierStates} to construct multiparticle Floquet-Bloch bands.
We consider an $N$-particle system with $L$ unit cells. 
For each cell, we focus on the case with three sublattices. The situations with other numbers of sublattices can be treated in a similar fashion.
To simplify the demonstration and without loss of generality, we require $N$ and $L$ to be coprime for $N>1$. Since $\lambda=2\pi/3$, the onsite potential term in $H_r$ is a periodic function of period $3$. Thus there are $3$ sublattices in each unit cell. We next define a co-translation operator $\widehat{T}_{3}$ that acts on Fock states as
\begin{align}
	&\widehat{T}_{3} \ket{n_1,n_2,\ldots, n_{3L}}\nonumber\\
	= &\ket{n_{3L-2},n_{3L-1},n_{3L}, n_1,\dots,n_{3(L-1)}},
\end{align}
i.e., $\widehat{T}_{3}$ translates the center-of-mass of particles over three sublattices to the left under the PBC. One can readily verify that $\widehat{T}_{3}$ reduces to the familiar translation operator in the one-particle case. For the rotated Hamiltonian in Eq.~(\ref{rotated}) and under the PBC, we have 
\begin{equation}
\widehat{T}_{3}H_{r}(t,\beta)\widehat{T}_{3}^{\dagger}=H_{r}(t,\beta). \label{commuteH}
\end{equation}

For two arbitrary Fock states $\ket{\mathbf{n}_{1}}$ and $\ket{\mathbf{n}_{2}}$, there may not exist a $j\in\mathbb{Z}$ such that $\ket{\mathbf{n}_{2}}=\widehat{T}_{3}^{j}\ket{\mathbf{n}_{1}}$. 
We define a set $S$ comprising of the so-called seed states~\cite{multiparticleWannierStates}. 
The elements of $S$ include all the Fock states such that any two of them cannot be transformed into each other by $\widehat{T}_{3}^{j}$ for any $j\in\mathbb{Z}$. Note that the set $S$ is in fact an equivalence relation. 
For example, for one-particle case, with a particle on either sublattice of the first unit cell, we get one $S$ comprised of $3$ seed states as $S=\{\ket{1},\ket{2},\ket{3}\}$.
For a general $N$, the dimension of Fock space is $D=\binom{3L+N-1}{N}$. 
Under the assumption that $L$ and $N$ are coprime, each seed state will go back to itself after being acted by $\widehat{T}_{3}$ over $L$ times. If $L$ and $N$ are not coprime, the number of actions of $\widehat{T}_{3}$ to recover a seed state is not unique, and we will not consider this case. With $L$ and $N$ being coprime, the number of seed states, i.e., the cardinality of set $S$ is $D_{S}=D/L$. Because of Eq.~(\ref{commuteH}), $\widehat{T}_{3}$ also commutes with $U(\beta)$ and they have common eigenstates $\ket{\psi}$, i.e.,
\begin{align}
    U(\beta)\ket{\psi}&=e^{i\epsilon}\ket{\psi},\\
    \widehat{T}_{3}^{-1}\ket{\psi}&=e^{i\phi}\ket{\psi}.
\end{align}
After some algebra (see Appendix \ref{derivation}), we find for any $\ket{\mathbf{m}}\in S$ that
\begin{equation}
    e^{i\epsilon}\braket{\mathbf{m}|\psi}=\sum_{\ket{\mathbf{n}}\in S}\bra{\mathbf{m}}U(\beta)\sum_{j=0}^{L-1}e^{ij\phi}\widehat{T}_{3}^{j}\ket{\mathbf{n}}\braket{\mathbf{n}|\psi}.\label{eigEqn}
\end{equation}
Therefore, we can reduce the full matrix $U(\beta)$ of dimension $D\times D$ to a reduced Floquet operator $\widetilde{U}(\beta,\phi)$ of dimension $D_{S}\times D_{S}$, whose matrix elements are 
\begin{equation}
    \bra{\mathbf{m}}U(\beta)\sum_{j=0}^{L-1}e^{ij\phi}\widehat{T}_{3}^{j}\ket{\mathbf{n}},\quad\ket{\mathbf{m}},\ket{\mathbf{n}}\in S. \label{reducedFloquet2}
\end{equation}
The Floquet states of $\widetilde{U}(\beta,\phi)$ are given by
\begin{equation}
    \ket{\widetilde{\psi}(\beta,\phi)}=[\braket{\mathbf{m}|\psi}]_{\ket{\mathbf{m}}\in S}^{\top}.
\end{equation}
Here $\top$ means transpose, so that $\ket{\widetilde{\psi}(\beta,\phi)}$ is a column vector. Note that $\ket{\phi,\mathbf{n}}=\frac{1}{\sqrt{L}}\sum_{j=0}^{L-1}e^{ij\phi}\widehat{T}_{3}^{j}\ket{\mathbf{n}}$ is just a Bloch eigenstate, and $\ket{\widetilde{\psi}(\beta,\phi)}=[\braket{\mathbf{m}|\psi}]_{\ket{\mathbf{m}}\in S}^{\top}$ is the eigenvector projected over seed states. By solving this eigenvalue problem, we get $D_{S}$ quasienergy bands. Note that there is a distinction between one-particle and multi-particle cases. For the one-particle case, the number of bands is always three, i.e., $D_{S}=3$, regardless of the value of $L$. We can use a splitting operator scheme to calculate the $3\times 3$ reduced Floquet operator $\widetilde{U}(\beta,\phi)$ for one-particle case, which is much faster than evaluating $\widetilde{U}(\beta,\phi)$ for the two-particle case (see the Appendix A of Ref.~\cite{kickedHarperModel} for more derivation details).

For the two-particle case, the value of $D_{S}$ changes with $L$. 
As long as $D_{S}\gg 1$, the exact number of bands $D_{S}$ does not generate a difference in the overall band structure. In the subsequent calculation, we take $L=21$. The number of bands is thus $D_{S}=96$ for $N=2$. We can then exponentiate the full $D\times D$ Hamiltonian, and multiply these exponentials according to their time ordering. For the one- or two-particle case, the exponential of Hamiltonian is manageable. If we have more particles, the computation will be more challenging as the dimension of the Hamiltonian increases exponentially and quickly becomes computationally prohibitive for exact diagonalization treatments. Note that our scheme of evaluating the reduced Floquet operator in Eq.~(\ref{reducedFloquet2}) is  equivalent to the Eq.~(3) in Ref.~\cite{multiparticleWannierStates}.

Meanwhile, our scheme has one computational advantage. That is, in order to get each matrix element of $\widetilde{U}$, we do not need to compute matrix-vector products. Instead, we only need to compute the inner product between vectors. Since $\bra{\mathbf{m}}U(\beta)$ is a row of $U(\beta)$, we only need to select this row to perform the inner product with $\sqrt{L}\ket{\phi,\mathbf{n}}$. Once we have obtained an eigenvector $\ket{\widetilde{\psi}(\beta,\phi)}$ over seed states, we can recover the full eigenstate through the relation
\begin{align}
    \ket{\psi(\beta,\phi)}&=\sum_{\ket{\mathbf{n}}\in S}\sum_{j=0}^{L-1}\widehat{T}_{3}^{j}\ket{\mathbf{n}}\bra{\mathbf{n}}\widehat{T}_{3}^{-j}\ket{\psi(\beta,\phi)}\nonumber\\
    {}&=\sum_{\ket{\mathbf{n}}\in S}\sum_{j=0}^{L-1}\widetilde{\psi}_{\mathbf{n}}e^{ij\phi}\widehat{T}_{3}^{j}\ket{\mathbf{n}}.
\end{align}

Our pumping scheme uses Wannier or Gaussian states as initial states. Their (unnormalized) explicit expressions are given by (i) Wannier state:
\begin{align}
    \ket{W(R_{0},\beta)}&=\sum_{\phi}e^{-i\phi R_{0}}\ket{\psi(\beta,\phi)}\nonumber\\
    {}&=\sum_{j=0}^{L-1}\sum_{\ket{\mathbf{n}}\in S}\sum_{\phi}\widetilde{\psi}_{\mathbf{n}}e^{i(j-R_{0})\phi}\widehat{T}_{3}^{j}\ket{\mathbf{n}},
\end{align}
and (ii) Gaussian state:
\begin{align}
    \ket{G(R_{0},\beta)}&=
    \sum_{j=0}^{L-1}\sum_{\ket{\mathbf{n}}\in S}\sum_{\phi}\widetilde{\psi}_{\mathbf{n}}e^{i\phi(j-R_{0})-\frac{1}{4\sigma^{2}}\phi^{2}}\widehat{T}_{3}^{j}\ket{\mathbf{n}}.
\end{align}
In the study of adiabatic pumping, we impose PBC for the rotated Hamiltonian $H_{r}$. We also obtain the Floquet bands with respect to the phase shift $\beta$ and center-of-mass quasimomentum $\phi\in[0,2\pi]$ by diagonalizing the Floquet operator under the PBC.

\section{Continuously driven AAH model with one boson}\label{Sec:boson1}

\begin{figure*}
	\begin{centering}
		\includegraphics[width=1\textwidth]{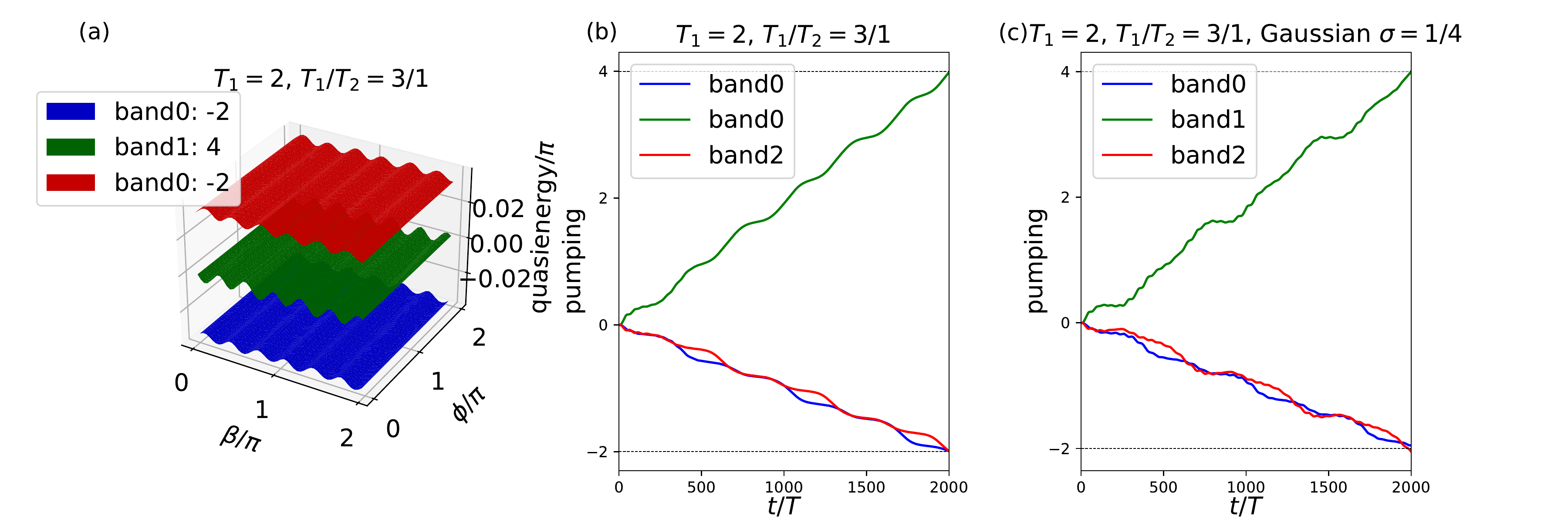}
		\par\end{centering}
	\caption{Floquet spectrum under the PBC, the pumping of Wannier and Gaussian initial states in real space for $T_{1}=2$, $a=3$ and $b=1$. (a) The three quasienergy bands with Chern numbers -2, 4, -2. (b) Pumping of Wannier states prepared in the three Floquet bands over an adiabatic cycle ($t=2000T$). (c) Pumping of Gaussian states initialized in the three Floquet bands over an adiabatic cycle ($t=2000T$). For each Wannier or Gaussian initial state, the shift of wavepacket center over an adiabatic cycle yields the Chern number of the corresponding Floquet band.\label{figs1pgaussian}}
\end{figure*}

We first demonstrate the Floquet band engineering in the single-particle case. In this case, we find topological transitions induced by changing the commensurate ratio $\omega_{F}/\Omega=a/b$ between the two driving frequencies. Moreover, the change of linear potential strength $\omega_{F}$ leads to new topological phases with large Chern numbers and many chiral edge states. We further employ the quantized adiabatic pumping of both Wannier and Gaussian states to reveal the rich Floquet topological properties induced by Bloch oscillations in the system from a dynamical perspective. Throughout this work, we work in dimensionless units and set the system parameters $J=V=2.5$.

\begin{figure*}
	\begin{centering}
		\includegraphics[width=1\textwidth]{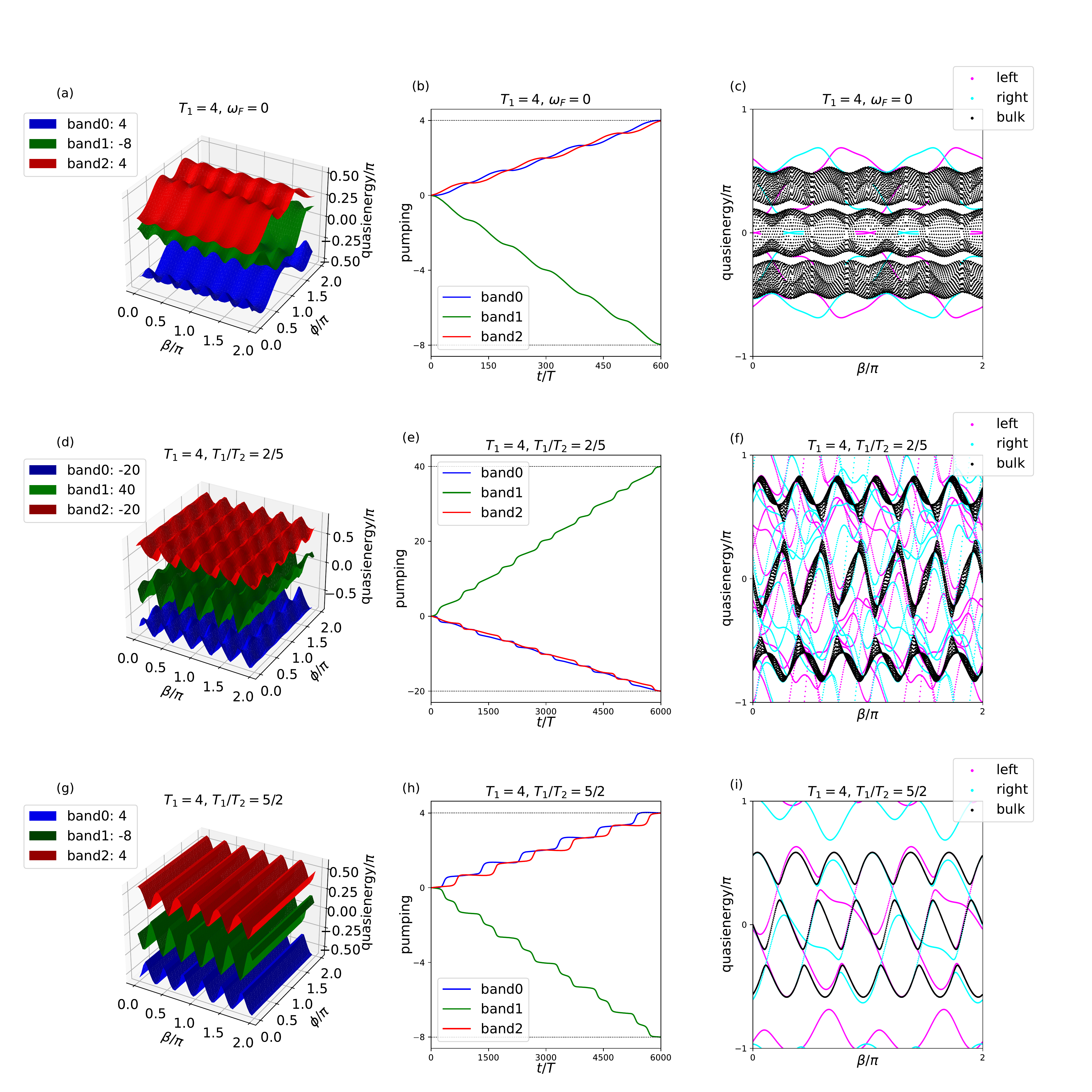}
		\par\end{centering}
	\caption{Floquet spectrum under the PBC, Wannier state pumping and Floquet spectrum under the OBC for $(T_{1}=4,\omega_{F}=0)$ in (a)--(c) and $T_{1}/T_{2}=(2/5, 5/2)$ in (d)--(i).}\label{figs1pT11Main}
\end{figure*}

We first present the Floquet bands under the PBC and the adiabatic pumping of Wannier and Gaussian states $\ket{W(R_{0},0)}$ and $\ket{G(R_{0},0)}$ in Figs.~\ref{figs1pgaussian} and \ref{figs1pT11Main}, where we take $T_{1}=2$, $a=3$ and $b=1$. We observe that the quasienergy bands are flat along the direction of quasimomentum $\phi$, as shown in Fig.~\ref{figs1pgaussian}(a) and Fig.~\ref{figs1pT11Main}(g). This holds true in other cases with $T_1/T_2>1$.
The flat dispersion is originated from Bloch oscillations induced by the linear potential. It could lead to the uniform sampling of initial states in momentum space, which is essential for the realization of quantized pumping for an arbitrary single-band state.
Regardless of the initial momentum distribution, each state is linearly swept in the momentum space multiple times in a single Floquet period when $T_1/T_2 > 1$, as shown in the Appendix~\ref{densityEvo} (see Fig.~\ref{fig:density_evolution}), leading to the flat band in the $\phi$-direction. The easiness of the flat-band generation is one of the benefits of employing the tilted potential, as it allows us to perform the Thouless pumping with arbitrary single-band initial states. To demonstrate the availability of quantized pumping with an arbitrary single-band initial state, we use the Gaussian state as an example. In Figs.~\ref{figs1pgaussian}(b) and \ref{figs1pgaussian}(c), we show the pumping of Wannier and Gaussian states prepared in the three Floquet bands over an adiabatic cycle consisting of $2000$ driving periods. For each case, we observe that the pumping is quantized and the net drift of wavepacket center  coincides with the Chern number of the corresponding Floquet band. Importantly, the quantized pumping of Gaussian initial state is made possible due to the flat quasienergy band along $\phi$-direction. Compared with the Wannier state, a Gaussian wavepacket is easier to prepare in experiments. Therefore, the linear potential provides us with a flexible route to achieve Thouless pumping and measure Chern numbers of Floquet bands in quantum simulators like ultracold atoms in optical lattices.

By changing the strength of the linear potential, the band topology of the system is modified. We consider a group of spectrum and pumping results with other frequency ratios including $\omega_{F}=0$, $T_{1}/T_{2}=2/5$ and $5/2$. The Floquet spectrum under the PBC without linear potential is shown in Fig.~\ref{figs1pT11Main}(a)~\cite{aspectsOf}. Here, we observe three dispersive bands with Chern numbers $4$, $-8$, $4$. In Fig.~\ref{figs1pT11Main}(b), the pumping of Wannier states indeed gives the Chern numbers of different Floquet bands. In Figs.~\ref{figs1pT11Main}(d) and \ref{figs1pT11Main}(e), we turn on the linear potential $\omega_{F}$ such that $T_{1}/T_{2}=2/5$. Interestingly, we see that the curvature of Floquet bands changes dramatically in response to the linear potential, and their Chern numbers could reach values as large as $-20$, $40$, $-20$. Therefore, we could get topological phases with huge Chern numbers by simply changing the value of linear potential strength $\omega_{F}$. Physically, the Floquet bands with large Chern numbers are originated from the interplay between the Bloch oscillations induced by linear potential and the external periodic driving field. In Fig.~\ref{figs1pT11Main}(g), we take $T_1/T_2$ as the inverse of $2/5$. Although the Floquet bands in this case have the same Chern numbers as those in Fig.~\ref{figs1pT11Main}(a), their local band curvatures are totally different. Fig.~\ref{figs1pT11Main}(g) has flat band along the $\phi$-direction, while along $\phi$ the band in Fig.~\ref{figs1pT11Main}(a) is curved. The pumping results of Wannier states in Fig.~\ref{figs1pT11Main}(h) are consistent with the Chern numbers of the corresponding Floquet bands. We present three more groups of Floquet bands and pumping results by adjusting the ratio of $\omega_{F}/\Omega$ in Fig.~\ref{figs1pT14Appendix}. They indeed show that we can generate rich Floquet topological phases with different frequency ratios $\omega_{F}/\Omega$. For completeness, we show the Floquet spectrum of the system under the OBC in Figs.~\ref{figs1pT11Main}(c), \ref{figs1pT11Main}(f) and \ref{figs1pT11Main}(i), and observe that the band Chern numbers correctly predict the number of chiral edge modes traversing the bulk gap at each edge.

Putting together, we conclude that the competition between the periodic driving $e^{\pm i\omega_{F}t}$ in the hopping terms (due to the linear potential) and the harmonic driving term $\cos(\Omega t)$ (in the onsite superlattice potential) determines the spectral topology and also has a clear impact on the wavepacket dynamics. The interplay between the two driving fields further induces rich topological structures in Floquet bands, leading to quantized adiabatic transport of wavepackets over a long spatial range in the lattice due to the large band Chern numbers. The unique feature of this system is that with the driving frequency $\Omega$ fixed, by applying a linear field $\sum_{j}\omega_{F}j \hat{n}_{j}$ that breaks the translational invariance of the lattice, the motion of the wavepacket can be strongly affected/tuned by Bloch oscillations and exhibits highly nontrivial transport signatures. 

Some necessary discussions are in order.  First, although the drift in the rotated frame of reference is the same as in the lab frame, there is also one subtlety.  That is, working with a rotated Hamiltonian via the aforementioned gauge transformation with translational invariance is valid only for states far from the boundary. For this reason, in actual simulations done in real space, we take a sufficiently long lattice and make sure that the wavepacket never reaches the boundary of the system. The translational invariance of the rotated Hamiltonian thus becomes valid, yielding the Bloch momentum $\phi$ to block diagonalize the Floquet operator and to compute the spectrum. Second, {although our work and Ref.~\cite{topologicalPumpingAssisted} have both demonstrated quantized adiabatic pumping under a linear potential, there are notable differences. In Ref.~\cite{topologicalPumpingAssisted}, the linear potential assists the uniform sampling in the Brillouin zone, and the quantization of pumping is independent of the strength of linear potential. By contrast, here the strength of the linear field is crucial as it controls the Floquet band topology, leading to rather different pumping results depending on the frequency of the Bloch oscillations induced by the linear field.}

\section{Continuously driven AAH model with two bosons}\label{Sec:boson2}

\begin{figure*}
	
	\begin{centering}
		\includegraphics[width=1\textwidth]{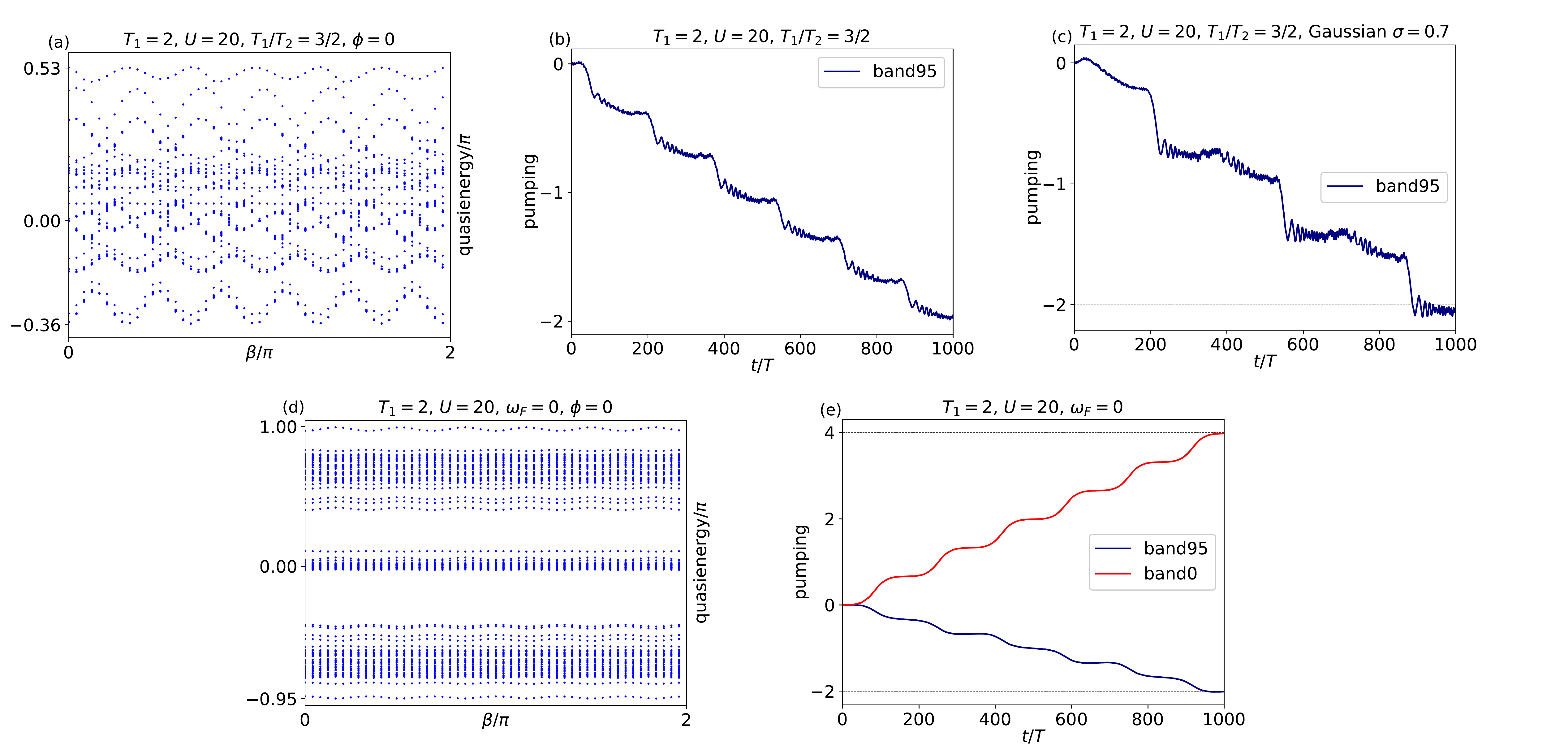}
		\par\end{centering}
	\caption{Floquet spectrum under the PBC and adiabatic pumping of two interacting bosons. The driving period $T_1=2$ and interaction strength $U=20$ are chosen for all panels. (a) Two-particle quasienergy spectrum for $T_{1}/T_{2}=3/2$ along $\phi=0$. (b) Pumping of a Wannier initial state for $T_{1}/T_{2}=3/2$. (c) Pumping of a Gaussian initial state for $T_{1}/T_{2}=3/2$. (d) Quasienergy spectrum for $\omega_{F}=0$ along $\phi=0$. (e) Pumping of two Wannier initial states for $\omega_{F}=0$.\label{figs2pT12U20Main}}
\end{figure*}

\begin{figure*}
	\begin{centering}
\includegraphics[width=1\textwidth]{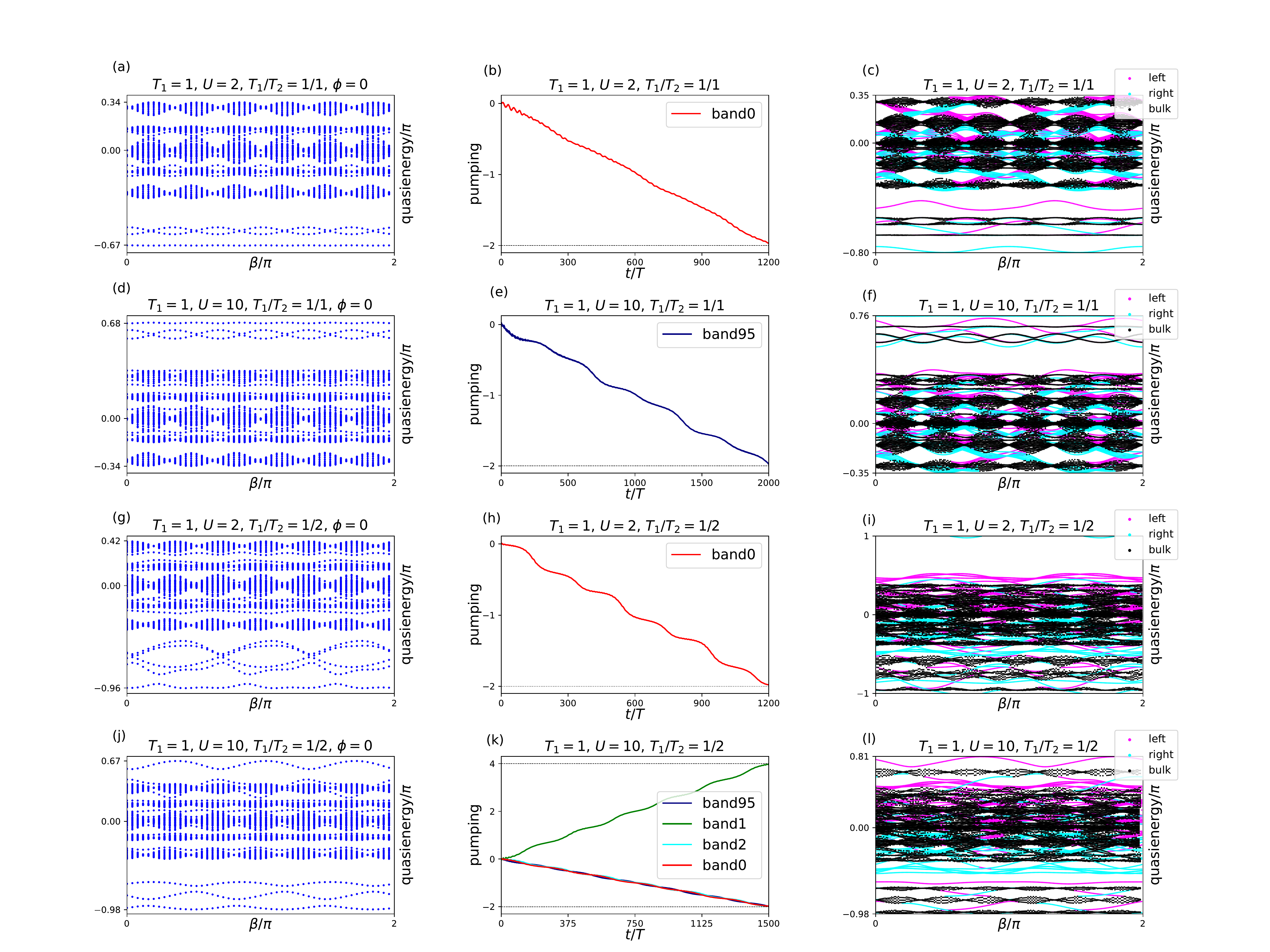} 
		\par\end{centering}
	\caption{Floquet spectrum under the PBC along $\phi=0$ in (a), (d), (g), (j), Wannier state pumping in (b), (e), (h), (k) and Floquet spectrum under the OBC in (c), (f), (i), (l) for interaction strengths $U=2,10$ and driving period ratios $T_{1}/T_{2}=1/1,1/2$.\label{figs2pT11Main}}
\end{figure*}

In this section, we study the Floquet spectrum and Thouless pumping of two interacting bosons in our system. We use the python package QuSpin~\cite{quspin} to treat the two-boson case, where exact diagonalization is applied. With two bosons, we have a total of $D_{S}=96$ bands under the PBC (still for a unit cell there are three sublattice sites), and it is tedious to visualize all Floquet bands using three-dimensional plots. Therefore, we focus on the Floquet bands obtained along a cut at $\phi=0$. The band structures obtained at other values of $\phi$ do not show noteworthy differences. We shall see that rich topological structures of Floquet bands, topological phase transitions and pumping dynamics can be generated by changing the ratio $T_{1}/T_{2}$ between two driving periods and the interaction strength $U$. The interplay among the periodic driving, interactions and Bloch oscillations thus induce nontrivial many-body Floquet topological phases.

The main results of this section are presented in Figs.~\ref{figs2pT12U20Main} and \ref{figs2pT11Main}. In Fig.~\ref{figs2pT12U20Main}(a), we show the two-particle Floquet spectrum under the PBC for $T_{1}=2$, $U=20$, and $T_{1}/T_{2}=3/2$. The highest Floquet band (denoted by band $95$) is found to be isolated from the rest of spectra. We hence choose to prepare the Wannier/Gaussian initial states on this band. Here Gaussian states refer to a Gaussian profile of the translational state of the center of mass of the bipartite system.  Results of our adiabatic pumping using Wannier and Gaussian states as the initials states are shown in Figs.~\ref{figs2pT12U20Main}(b) and \ref{figs2pT12U20Main}(c). The width of the Gaussian state in momentum space is set as $\sigma=0.7$. We find that the drift of wavepacket center over an adiabatic cycle is the same for the Gaussian and Wannier initial states. They both reproduce the Chern number $C=-2$ of band $95$. Therefore, we have successfully observed quantized pumping for two types of initial wavepackets. Of particular interest, the quantization of the pumping using only a two-particle Gaussian state is made possible by the linear potential induced Bloch oscillations as in the single-particle case. This indicates that the stage of initial state preparation in Thouless pumping can be greatly simplified even in many-body interacting cases thanks to the Bloch oscillations.  As a comparison, we show in Figs.~\ref{figs2pT12U20Main}(d) and \ref{figs2pT12U20Main}(e) the two-particle Floquet bands under the PBC without a linear potential, the pumping of Wannier states prepared on the lowest band (named band $0$) and the highest band (named band $95$) without the linear potential. The results show that the presence of a linear potential leads to quantized pumping at different integer values. As such, in the case of two interacting bosons, topological phase transitions can be also induced by turning on a linear potential $\omega_{F}$.

Next, we present four groups of results in Fig.~\ref{figs2pT11Main} for two-body Floquet spectrum under different boundary conditions, in connection with the quantized two-particle pumping. We set $T_{1}=1$ in all panels of Fig.~\ref{figs2pT11Main}, while choosing the interaction strengths to be $U=2,10$ and setting the ratios of $T_{1}/T_{2}$ as $1/1$ and $1/2$. One goal here is to examine the impact of interaction on the topological pumping. First, comparing the results presented in Figs.~\ref{figs2pT11Main}(a) and \ref{figs2pT11Main}(g), we see that the decreasing of the ratio $T_{1}/T_{2}$ from $1/1$ to $1/2$ for $U=2$ makes the lowest band more dispersive. Meanwhile, the number of groups of Floquet bands is kept to be the same as seven. On the other hand, by comparing Figs.~\ref{figs2pT11Main}(d) and \ref{figs2pT11Main}(j), we see that for $U=10$ the decreasing of ratio $T_{1}/T_{2}$ from $1/1$ to $1/2$ leads to more apparent differences. Both cases have the highest band (band $95$) as an isolated band, while three more isolated bands appear at the bottom in Fig.~\ref{figs2pT11Main}(j), as denoted by bands $0$, $1$ and $2$.  
The change of interaction strength $U$ also generates significant differences in the spectrum, topological and transport nature of the system. With the same ratio $T_{1}/T_{2}=1/1$, Figs.~\ref{figs2pT11Main}(a) and \ref{figs2pT11Main}(d) show the same number of seven groups of bands, but the isolated band goes from the bottom to the top. We can also observe that Fig.~\ref{figs2pT11Main}(a) is similar to the image being turned upside down of Fig.~\ref{figs2pT11Main}(d). Likewise, by comparing the results shown in Figs.~\ref{figs2pT11Main}(g) and \ref{figs2pT11Main}(j), we see that with the same ratio $T_{1}/T_{2}=1/2$, the number of isolated bands increases from one to four by increasing $U$ from $2$ to $10$, and the number of groups of bands increases from seven to nine. We conclude that not only the change of the ratio $T_{1}/T_{2}$, but also the change of the interaction strength $U$ will lead to topological phase transitions in the two-particle system. The Chern numbers in each case of Figs.~\ref{figs2pT11Main}(a), \ref{figs2pT11Main}(d), \ref{figs2pT11Main}(g), and \ref{figs2pT11Main}(j) are obtained by calculating Berry curvatures of Floquet bands under the PBC, and are validated by the quantized two-particle pumpings in Figs.~\ref{figs2pT11Main}(b), \ref{figs2pT11Main}(e), \ref{figs2pT11Main}(h), and \ref{figs2pT11Main}(k). Therefore, we provide concrete examples of topological phase transitions and quantized transport in two-particle Floquet systems that are induced by the interplay among a periodic driving field, interactions and Bloch oscillations.

In Appendix~\ref{moreExamples}, we have presented more computational examples. There it is seen that Floquet bands with very high Chern numbers can be obtained by simply adjusting the ratio of the Bloch oscillation frequency to that of the periodic driving. These additional results further strongly indicate that the introduction of a linear field to periodically driven lattices is a powerful means towards Floquet-band engineering.  

\section{Discussion and summary}\label{Sec:sum}
Experimentally, our model and the pumping dynamics may be realized in cold atom systems.
A 1D static AAH model can be realized by superimposing two optical lattices with different lattice constants, which result from standing waves created by retro-reflected single-frequency laser beams \cite{harperexperiment}. To realize the continuously driven AAH model, we may periodically modulate the strength of lattice lasers. The tunable linear field can be generated by a magnetic field gradient along the lattice direction \cite{blochexperiment}. Different magnetic field gradients lead to different band topology of the system, thus generating different quantized pumping results.

We have shown that a linear potential applied to a Floquet system can induce quasienergy bands with large Chern numbers and rich topological phase transitions. This is verified by calculating the topological invariants as well as adiabatic Thouless pumping of Wannier states. Moreover, we have shown that we can use Gaussian initial states on relatively flat bands to achieve quantized pumping, which is easier to implement in experiments. It is the Bloch oscillation that facilitates the uniform sweeping of eigenstates for such flat bands. Certainly, in Floquet systems, there is no longer uniform sampling of momentum by the Bloch oscillations, as evidenced by the existence of non-flat bands. In these cases, the linear potential can even induce new topological phases.   This is the main message from this work, and we hence expect that the introduction of a linear field can be generally useful to multi-color engineering of Floquet bands \cite{PhysRevB.100.235452}.   
For a two-particle interacting system,  not only the ratio of the two driving periods but also the interaction strength determines the topological phases and hence changes the result of Floquet Thouless pumping.
 In future work, we may consider situations with more bosons or interacting fermions~\cite{MuPRB2019}, in connection with many-body stark localization and quantum pre-thermalization.   

\begin{acknowledgments}
L.Z. is supported by the National Natural Science
Foundation of China (Grant No.~11905211), the Young
Talents Project at Ocean University of China (Grant 
No.~861801013196), and the Applied Research Project of Postdoctoral Fellows in Qingdao (Grant No.~861905040009). J.G. acknowledges support from Singapore National Research Foundation Grant No.~NRF-NRFI2017-04 (WBS No.~A-0004162-00-00).

The computational work for this article was fully performed on resources of the National Supercomputing Centre, Singapore (https://www.nscc.sg).
\end{acknowledgments}



\appendix
\section{Derivation of the eigenvalue problem}\label{derivation}
In this section, we give derivation details of Eq.~(\ref{eigEqn}) in the main text. For common eigenstates of the Floquet operator $U(\beta)$ and co-translation operator $\widehat{T}_{3}^{-1}$, we have
\begin{align}
    U(\beta)\ket{\psi}&=e^{i\epsilon}\ket{\psi},\\
    \widehat{T}_{3}^{-1}\ket{\psi}&=e^{i\phi}\ket{\psi}.
\end{align}
For the set $S$ of all seed states, the identity operator is
\begin{equation}
    \mathds{1}=\sum_{\ket{\mathbf{n}}\in S}\sum_{j=0}^{L-1}\widehat{T}_{3}^{j}\ket{\mathbf{n}}\bra{\mathbf{n}}\widehat{T}_{3}^{-j}.
\end{equation}
Then we obtain
\begin{align}
    e^{i\epsilon}\ket{\psi}&=U(\beta)\ket{\psi}\nonumber\\
    {}&=U(\beta)\sum_{\ket{\mathbf{n}}\in S}\sum_{j=0}^{L-1}\widehat{T}_{3}^{j}\ket{\mathbf{n}}\bra{\mathbf{n}}\widehat{T}_{3}^{-j}\ket{\psi}\nonumber\\
    {}&=\sum_{\ket{\mathbf{n}}\in S}U(\beta)\sum_{j=0}^{L-1}e^{ij\phi}\widehat{T}_{3}^{j}\ket{\mathbf{n}}\braket{\mathbf{n}|\psi}.
\end{align}
Multiplying $\ket{\mathbf{m}}\in S$ from the left, we get
\begin{equation}
    e^{i\epsilon}\braket{\mathbf{m}|\psi}=\sum_{\ket{\mathbf{n}}\in S}\bra{\mathbf{m}}U(\beta)\sum_{j=0}^{L-1}e^{ij\phi}\widehat{T}_{3}^{j}\ket{\mathbf{n}}\braket{\mathbf{n}|\psi}.
\end{equation}
this is Eq.~(\ref{eigEqn}) in the main text.

\section{Evolution of particle density in momentum space}
\label{densityEvo}
In this appendix, we provide details for the time evolution of the single-particle density distribution in the momentum space. It allows us to directly observe the effect of Bloch oscillations in Floquet systems. The density distribution is given by 
\begin{align}
	|\psi_k|^2 = |\phi_{0,k}|^2 + |\phi_{1,k}|^2 + |\phi_{2,k}|^2, 
\end{align}
where
\begin{align}
	\phi_{i,k} = \frac{1}{\sqrt{L}} \sum_{j = 1}^{L} e^{-i k j}\psi_{j,i}.
\end{align} 
In numerical calculations, we choose $L = 500$ as the number of unit cells. In a single Floquet period, the momentum is linearly swept through the Brillouin zone multiple times when $T_1/T_2 > 1$, leading to flat quasienergy bands along the direction of Bloch quasimomentum. 

\begin{figure}[ht]
		\centering
		\includegraphics[angle=0,width=0.49\textwidth]{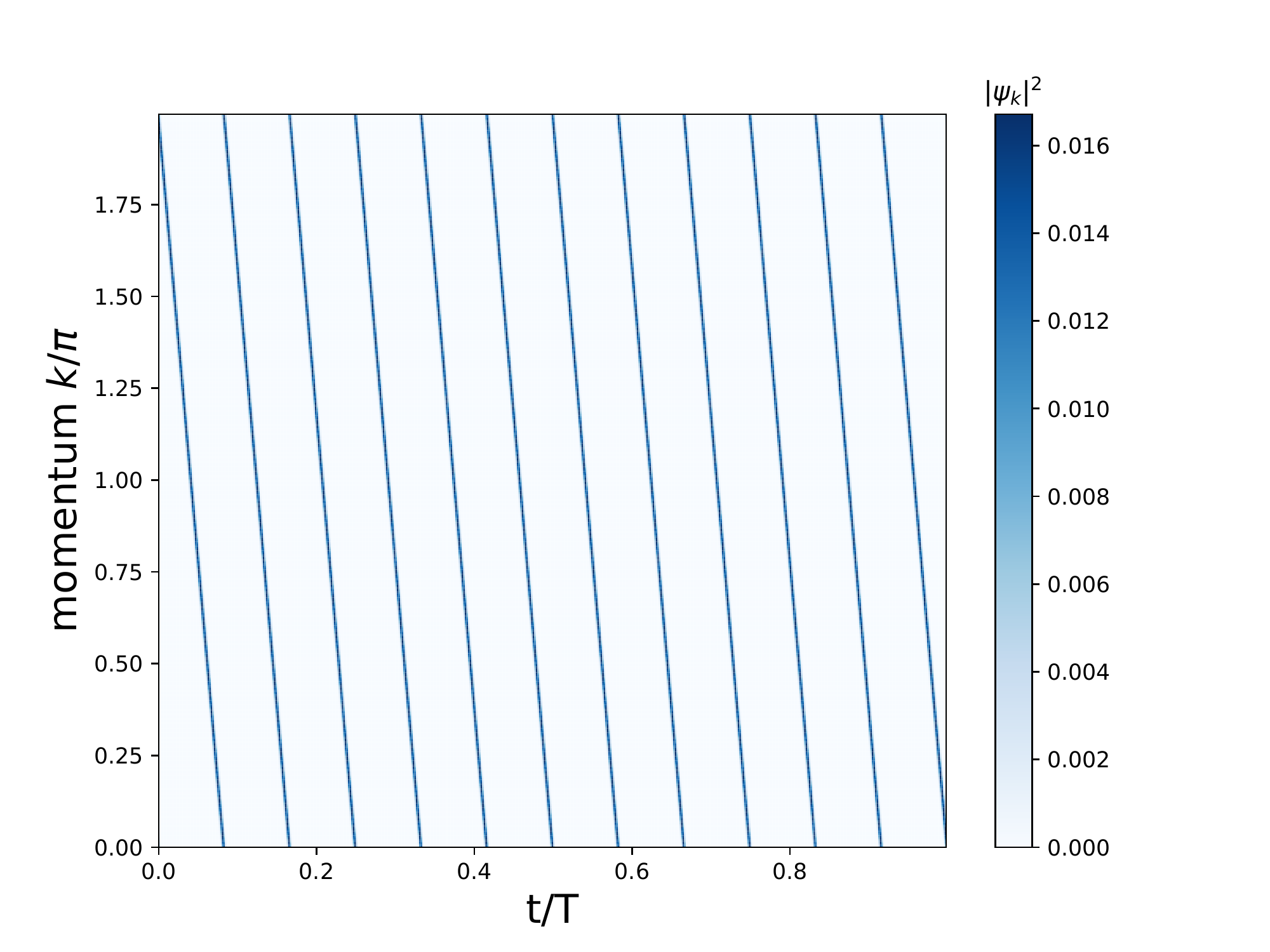}
		\caption{Time evolution of single-particle density distribution in the momentum space. The chosen parameters are $T = T_1 = 2, T_2 = 0.5, J = V = 2.5.$}\label{fig:density_evolution}
\end{figure}

\section{More examples for the single-particle case} \label{moreExamples}
\begin{figure*}[hbtp!]
		\begin{centering}
		\includegraphics[width=1\textwidth]{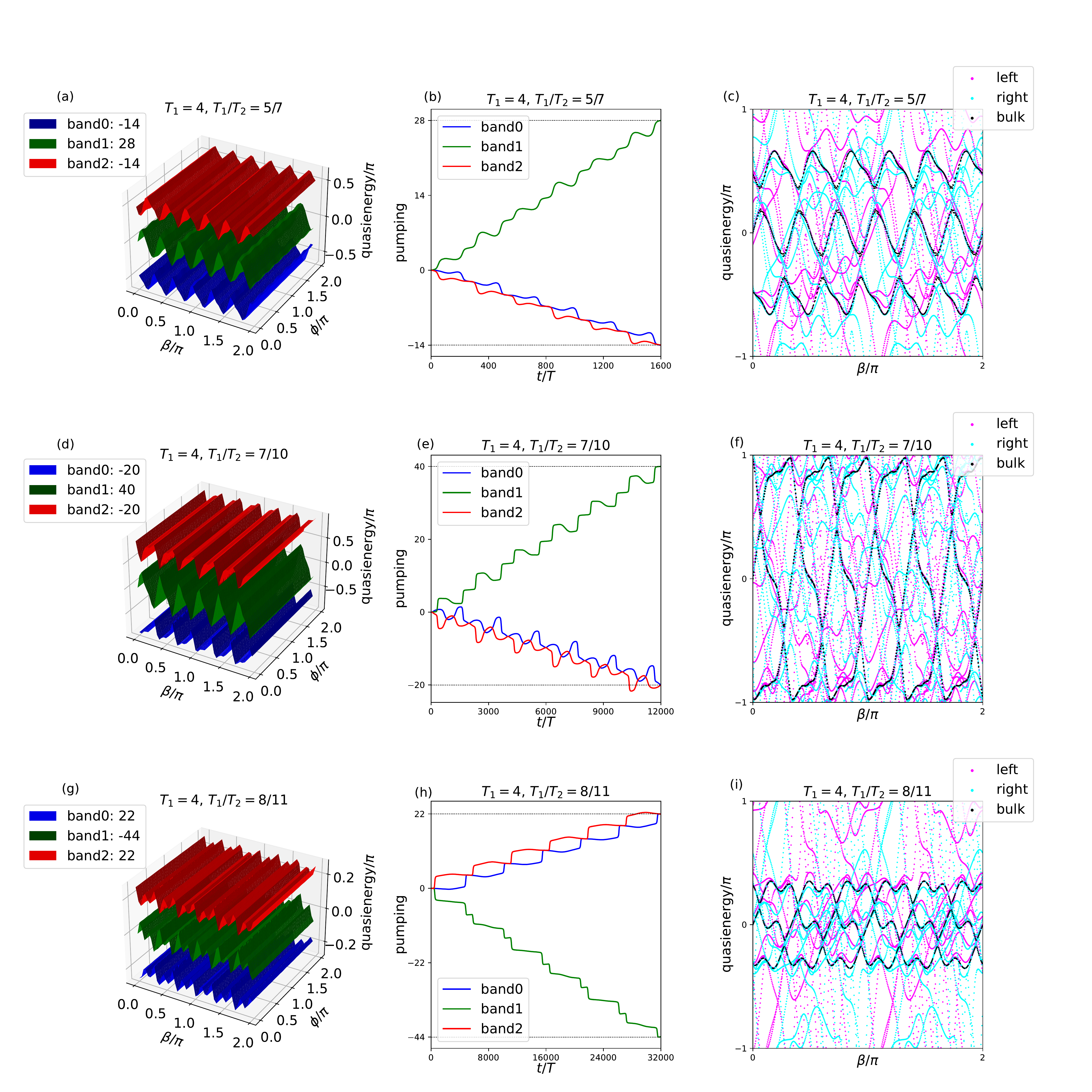}
		\par\end{centering}
	\caption{Floquet bands under the PBC, Wannier state pumping and Floquet bands under the OBC  for the single-particle case with $T_{1}=4$ and $T_{1}/T_{2}=5/7$ [in (a), (b), (c)], $7/10$ [in (d), (e), (f)], $8/11$ [in (g), (h), (i)], respectively.\label{figs1pT14Appendix}}
\end{figure*}
In this appendix, we present $3$ more examples of Floquet quasienergy spectrum and Thouless pumping in Fig.~\ref{figs1pT14Appendix}. We can see that by changing the ratio of $T_{1}/T_{2}$, one can easily obtain topological phases with very high Chern numbers.  It is also observed that the actual Chern numbers obtained can be rather sensitive to the strength of the applied linear field. This provides us with more flexibility to control topological phase transitions in Floquet systems.

\bibliography{apssamp}

\end{document}